# Self-stimulated Emission of Undulator Radiation

E.G.Bessonov†, M.V.Gorbunkov†, A.A.Mikhailichenko††, A.L.Osipov†

†Lebedev Phys. Inst. RAS, Moscow, Russia, ††Cornell University, LEPP, Ithaca, NY 14853, U.S.A.

*Abstract.* We attract attention that interaction of particle in downstream undulator with its own wavelet emitted in upstream undulator could be as strong as with the frictional field in undulator itself. This phenomenon could be used for enhancement of signal from pickup undulators in optical stochastic cooling methods as well as for increase of damping.

Particle passed an undulator emits undulator radiation wavelet (URW) which length is $M\lambda_1$ where $M$ is the number of undulator periods, $\lambda_1$ – is the wavelength of first harmonic. In system of $N$ identical undulators located along straight line the particle radiates train of wavelets with separation $l$; both $l$ and $\lambda_1$ defined by Doppler effect, by angle $\theta$ between instant velocity and direction to observer, by distance between undulators $l_0$, by period of undulator $\lambda_u$ and by relativistic factor $\gamma >> 1$. In straight forward direction $\theta = 0$ they are $l = l_0 / 2\gamma^2$ , $\lambda_1 = \lambda_u / 2\gamma^2$. Energy radiated by particle in system of $N$ undulators is $N$ times bigger than the one radiated in just one undulator. Spectrum of radiation emitted in arbitrary direction also changes: appears line-type spectrum. Integrated spectrum changes not much, see [1].

In this publication we suggest to increase the loss rate in system of $N$ undulators by introduction of controlled delays in motion of particles relative to the URW between undulators, Fig.1.

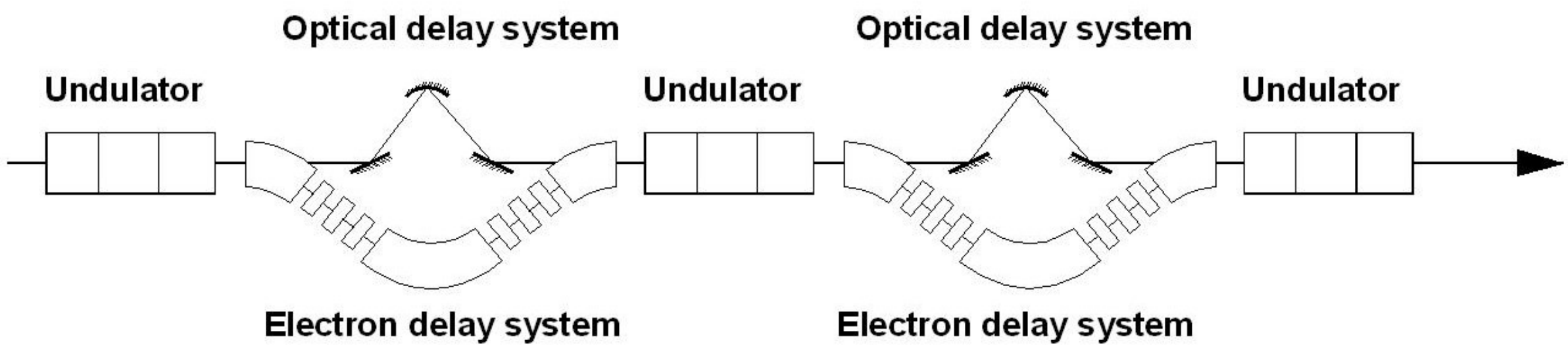

Figure 1: Scheme of installation.

Delays chosen so that particle enters the following undulator together with the front edge of URW emitted in anterior undulators in decelerating phase. In this case the particle will experience deceleration in its self field generated by its instant motion in a field of undulator (friction force generated by spontantenous incoherent radiation) as well as in the field of URW from anterior undulators (induced radiation in field of co-propagating electromagnetic wave). Under such condition occurs superposition of wavelets which yield the electric field grows ~$N$ so the energy emitted grows ~$N^2$.

To be effective and optimal, this system must use appropriate focusing elements such as lenses and/or focusing mirrors, see the scheme of installation on Figs. 1, 2. Mirrors and lenses must form crossover with the Rayleigh length of the order of the length of undulator $Z_R \cong M\lambda_u / 2$ [2].

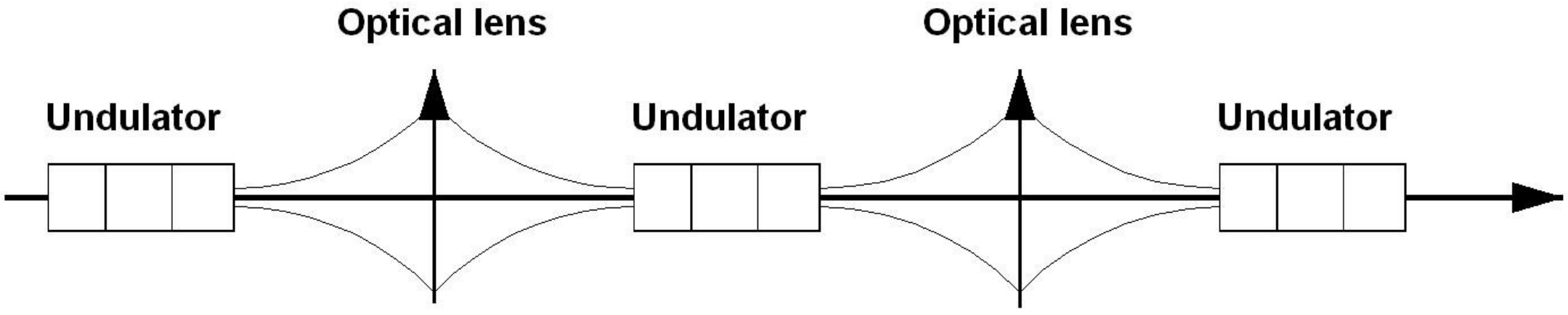

Figure 2: Equivalent optical scheme.

The scheme of installation suggested could be used effectively in different methods of optical cooling (OC) of particles in damping rings [2]-[5]. So this installation can serve as effective pick-up undulator. According to OC principle the optical parametric amplifier(s), controllable screens [2] and kicker undulators could be located in the subsequent straight sections.

We would like to remind here that for any method of Optical Stochastic Cooling it is important to inject in spectral bandwidth $\Delta\lambda / \lambda \sim 1/2M$ and in angles $\Delta\theta \sim 1/\gamma\sqrt{M}$ as many photons as possible. This number does not depend on the length of undulator [2]. So usage of three pickup undulators is 3 times more effective in the emitted field strengths and 9 times more effective in the emitted energy, than just single pickup undulator and so on. It means that usage of three pickup undulators and singe kicker one in the schemes of OSC is 3 times more effective for damping time, than just single pickup and single kicker undulator. Usage of three pickup undulators and three kicker undulator is 9 times more effective, than just single pickup and single kicker undulator and so on. So the effectiveness of the pickup and kicker systems consisting of $N$ undulators each is proportional to $N^2$.

We considered here the case when optical delays tuned so that wavelets emitted by particle are congruent and particle always stays in decelerating phase. To be so the beam delay system must be isochronous for all particles in the beam.

There is a possibility for another scheme with self-stimulated undulator radiation. This scheme uses isochronous storage ring with undulator installed in one straight section. Mirrors installed at both sides of undulator set an optical cavity so that period of oscillation of wavelet in optical cavity coincides with period of revolution of particles in storage ring. In that case the wavelets will be accumulated in optical cavity superimposed one by another with the accuracy $<< \lambda_1$. This scheme is typical for FEL, but the difference is that the storage ring is isochronous and that the motion of the wavelet and the bunch are synchronized one with another. In this case there is no coherence in radiation among different particles in the bunch (as the particles are not grouped in microbunches with longitudinal dimensions $\sigma_{||} << \lambda_1$ separated by distances which are integer of $\lambda_1$), but stimulated processes are going in their own fields of URWs emitted in undulator in earlier times.

All properties of spontantenous incoherent radiation emitted by particles in this case are not changed, except intensity, which becomes higher now in $Q$ times, where $Q$ is the quality factor of optical cavity. If however, conditions of synchronicity are broken weakly so the wavelets emitted at each pass through undulator are shifted by $\sim \lambda_1 \div M\lambda_1$, then properties of radiation might be different now (intensity will drop, but monochromaticity will be enhanced). If isochronicity satisfied for particles in some narrow diapason of angles and energy, then this will narrow angular divergence and spectrum of undulator radiation at the exit of optical cavity. Strong dependence of intensity of undulator radiation on energy may change the cooling rate of particles in storage ring [6].

This phenomenon of self-stimulated emission in undulator can be used for tuning optical system of any optical stochastic cooling schemes. This correspond operation of system with optical amplifier turned off and the optical delay shifted by $\sim \lambda/2$ with respect to the optimal cooling phase. So some tuning could be done without optical amplifier at all, if someone just registering intensity of forward radiation after kicker undulator. Then just by shifting optical delay back by half wavelength and tuning on optical amplifier, the system will be set to optimal phasing.

We would like to mention that in parametric FEL with mirrors [7], [8] (stimulated super-radiant emission in pre-bunched Free-Electron Laser) the process of radiation is similar to described in our paper. However there *new* portions of particles, bunched into small-size, passing through undulator with the same periodicity.

This work was supported in part by RFBR under Grant No 09-02-00638a.